# AI for Proactive Mental Health:

## A Multi-Institutional, Longitudinal, Randomized Controlled Trial

Julie Y. A. Cachia[1], Xuan Zhao[1,2], John Hunter[3],

Delancey Wu[4], Eta Lin[5], Julian De Freitas[6]

[1] Flourish Science Inc., Palo Alto, CA 94304

[2] Stanford University, Department of Psychology, Stanford, CA 94305

[3] Chapman University, Crean College of Health and Behavioral Sciences, Orange, CA 92866

[4] University of Washington, Department of Psychology, Seattle, WA 98195

[5] Foothill College, Psychology Department, Los Altos Hills, CA 94022

[6] Harvard University, Harvard Business School, Boston, MA 02163

**Corresponding Authors:** Julian De Freitas (jdefreitas@hbs.edu), Julie Y. A. Cachia (julie@myflourish.ai)

**Abstract**

Young adults today face unprecedented mental health challenges, yet many hesitate to seek support due to barriers such as accessibility, stigma, and time constraints. Bite-sized well-being interventions offer a promising solution to preventing mental distress before it escalates to clinical levels, but have not yet been delivered through personalized, interactive, and scalable technology. We conducted the first multi-institutional, longitudinal, preregistered randomized controlled trial of a generative AI-powered mobile app ("Flourish") designed to address this gap. Over six weeks in Fall 2024, 486 undergraduate students from three U.S. institutions were randomized to receive app access or waitlist control. Participants in the treatment condition reported significantly greater positive affect, resilience, and social well-being (i.e., increased belonging, closeness to community, and reduced loneliness) and were buffered against declines in mindfulness and flourishing. These findings suggest that, with purposeful and ethical design, generative AI can deliver proactive, population-level well-being interventions that produce measurable benefits.

Over the past decade, global rates of anxiety, depression, and loneliness have risen to unprecedent levels among young adults. In a nationally representative survey of over 80,000 students across 135 U.S. higher education institutions in 2024, 76% of respondents reported they currently "needed help for emotional or mental health problems," yet only 38% received support from a health professional within the past year, and 11% reported having considered suicide in the past year[1]. Structural and psychological barriers further discourage help-seeking, even among those who recognize their need. Lack of time, limited access, stigma, and confidentiality concerns remain among the leading deterrents[1,2]. Although only 7% of college students report that they would judge a peer for seeking help, 41% believe others would judge them, highlighting the powerful influence of perceived stigma[1]. Nationally, 62% of Americans needing care cite shame or stigma as obstacles[3], which is linked to reduced engagement with professional support[2].

In response to the growing demand, society has increasingly turned to scalable solutions such as digital mental health platforms[4] and teletherapy[5]. Advances in Artificial Intelligence (AI) have further demonstrated the potential for personalized, interactive, and always-available support at scale. By leveraging AI to replicate traditional clinical practices, chatbots can mirror therapist behaviors[6], deliver structured interventions like cognitive reframing therapy[7], and provide targeted feedback to peer counselors[8]. Consequently, a wave of AI-powered mental health chatbots have entered the market. While earlier systems were constrained by rule-based decision trees and limited input flexibility (e.g., Woebot[9], Tess[10], Wysa[11], Youper[12]), recent advances in large language models (LLMs) are beginning to enable more natural, adaptive, and personalized interactions for mental health treatment (e.g., Therabot[13]).

Although these tools show great promise in terms of accessibility and scalability, they often lack the nuance and empathy of human therapists[14,15]. Moreover, most reflect a deficit-based approach to mental health, which centers on identifying, diagnosing, and treating psychological symptoms after they have emerged[16]. This model typically emphasizes clinical markers such as depression, anxiety, and stress, treated with interventions like cognitive behavioral therapy (CBT)[17] or crisis counseling[18]. While these interventions are undoubtedly valuable, they are inherently reactive, leading many people to delay engagement until they reach a breaking point. They may also inadvertently reinforce the view that mental health support is a last resort rather than a foundational component of everyday well-being.

In light of the limitations of a deficits-based model, scholars have advocated for a proactive, "strengths-based" approach that emphasizes cultivating well-being, resilience, connection, and purpose rather than merely treating symptoms of clinical problems[19,20]. This perspective draws from positive psychology, public health, and preventative medicine, proposing that people may be more resilient to mental health challenges if they are equipped beforehand with a strong foundation of personal and social strengths[21].

Compared to focusing on symptom reduction, strengths-based interventions present two unique advantages: First, they can help foster positive emotions, belonging, resilience, mindfulness and overall well-being—factors that buffer against mental distress and promote flourishing. Second, by cultivating positive emotions, social connection, or resilience, they can make participation feel less like a response to deficiency and more like an act of personal growth. As a result, strengths-based initiatives may also sidestep the stigma associated with traditional deficits-based approaches and enjoy broader appeal and rates of adoption. Yet, so far, strengths-based approaches have predominantly been disseminated through static formats like books[22–24], podcasts, online courses, or workshop events. While these formats have successfully introduced the science of well-being to tens of millions of people, delivering the same content in a fixed, one-size-fits-all manner may not meet every user's mental state or needs. To date, few, if any, attempts have been made to integrate AI to deliver personalized, context-sensitive, just-in-time, strengths-based support. While one early finding suggests that structured chatbot dialogues can yield short-term well-being gains[25], these effects were measured at a single timepoint in a lab setting, underscoring the need for longitudinal studies in real field settings.

To this end, we employ Flourish (https://www.myflourish.ai/), a mobile app launched in 2024 that integrates generative AI with decades of research in well-being science to deliver personalized, gamified, strengths-based mental health support. At the centre of the app is Sunnie, an AI-powered well-being chatbot that engages users in brief, emotionally intelligent conversations and recommends relevant knowledge, insights, and science-based coping strategies. Unlike static tools, Sunnie uses a sophisticated workflow based on a variety of large language models (LLMs) to adapt in real time to users' mood, goals, and behavior, adjusting its tone, content, and suggestions to meet users where they are.

The Flourish app is designed to promote well-being through an intuitive behavior loop (Figure 1). The loop begins with a brief emotion check-in, followed by real-time support and feedback through multi-turn conversations with Sunnie that focus on building emotional awareness and learning relevant psychology knowledge. Next, Sunnie provides science-based activities recommendations—such as journaling for emotion regulation[26–28], positive psychology exercises (e.g., three good things[29]), social connection activities (e.g., gifting a compliment[30,31], meaningful conversations[32,33]), and mindfulness techniques (e.g., nature walks[34–36]). Then, each interaction is positively reinforced with rewards (e.g., AI-generated badges, streaks), and together, they contribute to an AI-generated weekly report, which synthesizes mood trends to help users reflect on their progress. In doing so, Flourish creates a closed, personalized feedback loop—moving from receiving support, to taking action, to gaining insight, and then back to renewed support interactions. This adaptive loop is informed by what we call the STAR framework for beneficial AI use—Science-based, Timely, Action-oriented (activating positive behavior change, rather than only talking about thoughts and feelings), and Real-life-focused (encouraging connections with other people and communities, rather than only with AI).

To evaluate whether this well-being intervention could serve as a scalable solution for college mental health, we conducted a six-week randomized controlled trial across three U.S. campuses (two four-year universities and one two-year community college). College students ($n = 486$) were randomly assigned to either a treatment condition (with access to the Flourish app and instructions to engage at least twice weekly) or to a control group (with existing school supports but no Flourish app access, hence maintaining status quo). This light-touch design aligns with the demands of busy student life and spans a key segment of the academic term, capturing natural stressors like midterms and deadlines.

Consistent with our strengths-based framework, we measured key outcomes related to emotional, social, and overall well-being—rather than symptom reduction alone. All psychological outcomes were assessed using validated psychological instruments (see Methods). Students completed brief online surveys (5–15 minutes each) across four waves (i.e., Week 0, 2, 4, 6). In particular, we sought to understand whether the treatment condition delivered benefits via actively boosting well-being, buffering against declines in well-being, or both simultaneously. Boosting and buffering produce patterns that are conceptually distinct: boosting reflects upward improvements in well-being relative to baseline, whereas buffering reflects protection from decline relative to a worsening control group.

**Fig. 1.** A high-level summary of the behavioral loop of the App Intervention.

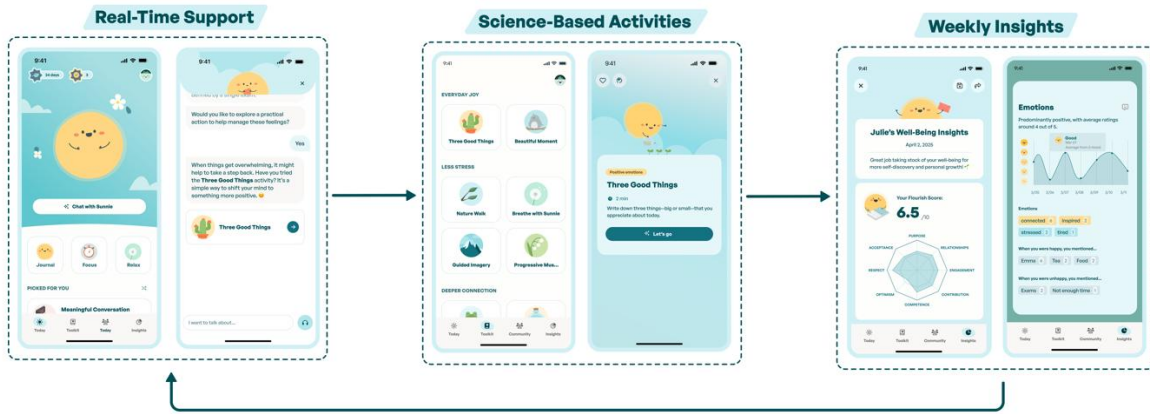

*Note.* Participants receive real-time support through interactive check-ins and conversations with Sunnie, the AI well-being coach and companion. These interactions lead into brief, interactive, and personalized science-based well-being activities. Weekly insights then summarize social and emotional patterns, helping users reflect and plan with clarity. See Methods and Supplementary Materials, Fig. S1 for a summary of the core generative AI components and Fig. S2 for a walkthrough of a typical daily user experience.

## Results

On average, participants used the app 3.49 days per week ($SD = 3.46$), exceeding the twice-per-week requirement and reflecting the intervention's intrinsic appeal. To test whether interacting with the Flourish app improved mental health outcomes, we ran longitudinal mixed-effects models using an intent-to-treat approach, including all participants regardless of engagement level. Each outcome was regressed on the interaction between Condition (treatment or control) and Time (Week 2, 4, or 6, respectively), with random intercepts included for both participant and institution. We expected to see significant Condition × Time interaction effects, in which the treatment condition elicited increasingly improved outcomes compared to control as time progressed. We summarize these results for each outcome variable below and in Figure 2 and Table 1 (for full details, see Supplementary Materials, Section 4; descriptive means and standard deviations are provided in Extended Data Table 3). Additional analyses examining whether intervention effects varied by demographics or baseline mental health revealed no statistically significant moderation effects after correcting for multiple comparisons (see Supplementary Materials, Section 6).

**Fig. 2**. Changes in emotional, social, and overall well-being outcomes as a function of Condition and Time.

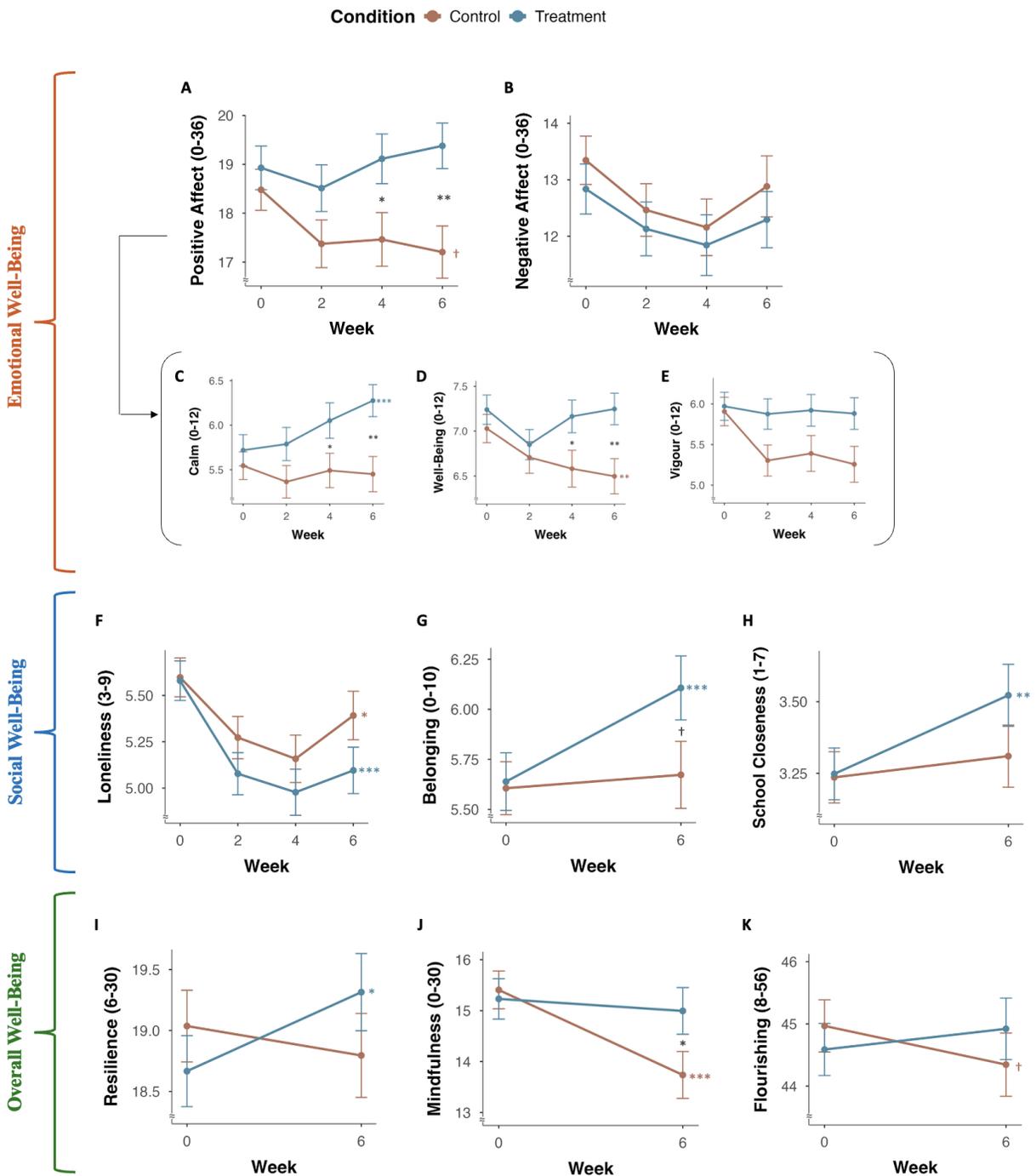

*Note.* Effects of condition on dependent variables across time. Error bars reflect the standard error of the mean. Significance symbols in colour represent the condition-specific time effects, whereas significance symbols in black represent the between-condition effects at a given timepoint. † *p* < .10 (marginal), **p* < .05, ***p* < .01, ****p* < .001.

***Emotional Well-Being***. Emotional well-being was measured by using the Subcomponents of Affect Scale[37] which assesses positive affect (including the subcomponents of calm, well-being, and vigour, averaged into a single construct of positive affect) and negative affect (including the subcomponents of depression, anxiety, and anger, averaged into a single construct of negative affect). Positive affect showed a significant Condition × Time interaction ($b = 0.22$, $SE = 0.10$, $t(1151.16) = 2.33$, $p = .020$, $β = 0.04$): by Weeks 4 and 6, the treatment group reported higher levels of positive affect than the control group (Week 4: $b = 0.82$, $SE = 0.37$, $t(352) = 2.21$, $p = .028$, $d = 0.23$; Week 6: $b = 1.09$, $SE = 0.35$, $t(346) = 3.07$, $p = .002$, $d = 0.33$), despite starting at comparable levels.

While the composite positive affect score showed no within-group effects, examining the subcomponents revealed distinct patterns: Participants in the treatment group showed a significant increase in "calm" (the low-arousal positive affect subscale), indicating a boosting effect ($b = 0.10$, $SE = 0.04$, $t(1173.98) = 2.49$, $p = .013$, $β = 0.04$); were protected against declines in "well-being" (the mid-arousal positive affect subscale) seen in the control group, consistent with a buffering effect ($b = 0.08$, $SE = 0.04$, $t(1162.58) = 2.25$, $p = .025$, $β = 0.04$); while "vigour" (the high-arousal positive affect subscale) showed no significant change in either group ($b = 0.05$, $SE = 0.04$, $t(1156.75) = 1.25$, $p = .212$). These findings suggest that the intervention supported emotional well-being by helping students boost lower-arousal positive states and buffer against declines in mid-arousal positive states (for detailed results, see Supplementary Materials, Section 5).

In contrast to positive affect, the Condition × Time interaction effect was non-significant on both the aggregated negative affect ($b = 0.0008$, $SE = 0.10$, $t(1151.68) = 0.08$, $p = .933$), as well as its three subcomponents ($p$'s > .516). This suggests that the current intervention may support emotional well-being by operating primarily on positive affect.

***Social Well-Being.*** We observed a significant Condition × Time interaction for loneliness ($b = -0.05$, $SE = 0.02$, $t(1158.52) = -2.08$, $p = .038$, $β = -0.03$) and belonging ($b = 0.06$, $SE = 0.03$, $t(370.89) = 2.18$, $p = .030$, $β = 0.04$), as well as a marginal interaction for closeness to the university ($b = 0.04$, $SE = 0.02$, $t(391.37) = 1.82$, $p = .070$, $β = 0.04$). These patterns suggest that the intervention primarily boosted social well-being rather than buffered against decline: While both treatment and control groups showed reduced loneliness over the semester—consistent with increasing social integration—the decline was significantly steeper for the treatment group, suggesting that the intervention boosted this loneliness decline. We see evidence for similar boosting effects for belonging and perceived closeness to the university, since these outcomes improved in the treatment groups while staying relatively flat in the control group. While prior research has shown that talking to chatbots can temporarily alleviate loneliness in a laboratory setting,[38] we suspect that the present results were further driven by the intervention's focus on taking real-world actions to connect with others (e.g., expressing gratitude), thereby strengthening belonging and perceived closeness to the campus community as well.

***Overall Well-Being.*** We observed a significant Condition × Time interaction for resilience ($b = 0.11$, $SE = 0.06$, $t(373.54) = 1.99$, $p = .047$, $β = 0.04$), mindfulness ($b = 0.19$, $SE = 0.09$, $t(386.69) = 2.02$, $p = .045$, $β = 0.05$), and flourishing ($b = 0.19$, $SE = 0.09$, $t(368.16) = 2.15$, $p = .032$, $β = 0.04$). Patterns for each measure were consistent with both a boosting and buffering effect of the intervention. Resilience increased over time in the treatment group but stayed steady in the control group, suggesting that the intervention boosted resilience. On the other hand, mindfulness declined in the control group but remained stable in the treatment group, while flourishing—a broad index of psychological well-being encompassing meaning, engagement, relationships, and self-acceptance—declined marginally in the control group but remained stable in the treatment group.

***Other Outcomes.*** Clinical outcomes—including depression, anxiety, and stress—as well as academic self-efficacy, did not show statistically significant changes ($p \geq .462$), although effects were generally in the expected direction. These findings suggest that, given the non-clinical nature of the sample and the current setup—including a relatively short duration with light app usage (approximately 5-20 minutes per week)—the intervention's immediate impact may be more pronounced for well-being outcomes than for clinical or academic measures (for detailed results, see Supplement Section 3).

**Table 1**

*Summary of Condition × Time Interaction Results*

| Outcome Variable | Condition × Time ($b$) | SE | df | t | $p$-value | $\beta$ |
|---|---|---|---|---|---|---|
| Positive Affect | 0.22 | 0.10 | 1153.08 | 2.33 | .020 | 0.04 |
| *Calm* | 0.10 | 0.04 | 1173.98 | 2.49 | .013 | 0.04 |
| *Well-Being* | 0.08 | 0.04 | 1162.58 | 2.25 | .025 | 0.04 |
| *Vigour* | 0.05 | 0.04 | 1156.75 | 1.25 | .212 | – |
| Negative Affect | 0.008 | 0.10 | 1151.68 | 0.08 | .933 | – |
| Loneliness | -0.05 | 0.02 | 1162.49 | -2.06 | .040 | -0.03 |
| Belonging | 0.06 | 0.03 | 371.49 | 2.15 | .032 | 0.04 |
| Closeness to School | 0.04 | 0.02 | 392.09 | 1.80 | .072 | 0.04 |
| Resilience | 0.12 | 0.06 | 374.67 | 2.11 | .036 | 0.04 |
| Mindfulness | 0.20 | 0.09 | 387.48 | 2.10 | .037 | 0.05 |
| Flourishing | 0.19 | 0.09 | 368.66 | 2.18 | .030 | 0.04 |

*Note.* Coefficients ($b$) reflect the Condition × Time interaction term for linear mixed-effects model.

## Discussion

AI chatbots are rapidly becoming everyday tools for emotional support, especially among young people[39]. Yet most lack evidence-based design to promote well-being, and concerning reports have emerged about their potential adverse effects on mental health[40]. Rigorous research is therefore needed to examine whether—and how—a generative AI system intentionally built to deliver science-based, strengths-focused support can improve mental health and well-being. Filling this gap, the current research offers the first preliminary evidence that a strengths-based, AI-powered intervention can offer personalized well-being benefits at scale. Specifically, interacting with the Flourish app improved positive affect, social well-being (increased belonging and closeness to community, reduced loneliness) and buffered against declines in mindfulness and flourishing across the semester. Effects on negative affect were non-significant, although participants reported higher resilience, suggesting an enhanced perceived capacity to cope with stressors.

At the same time, the absence of effects on depression, anxiety, or stress likely reflects both the nonclinical baseline of participants and the app's preventive, strengths-based focus. This intervention may therefore be best positioned as a proactive, scalable, well-being tool rather than a replacement for clinical treatment. Future work should also explore whether this intervention could help during clinical treatment—as in the form of hybrid intervention models that combine the AI agent with human therapy—or after treatment, to maintain the benefits of clinical treatment and prevent relapse.

One important limitation of the current research is that we adopted a treatment-versus-waitlisted-control design to measure improvement from status quo, rather than including an active control group. Accordingly, we cannot rule out that some effects may stem from general engagement (e.g., perceived support or demand characteristics) rather than the specific content of the intervention. However, qualitative evidence suggests meaningful engagement with and benefit from the app. Two independent coders conducted a content analysis of participants' open-ended responses after Week 6. To the question, "What was it like using the Flourish app?", many participants spontaneously described improvements in overall well-being (58%; $\kappa = .63$, agreement = 88%), emotional well-being (24%; $\kappa = .82$, agreement = 93%), and social well-being (10%; $\kappa = .79$, agreement = 95%), corroborating the quantitative results. Some even attributed the benefits to specific app features such as the chatbot (22%; $\kappa = .87$, agreement = 95%), activities (8%; $\kappa = .78$, agreement = 96%), or weekly insights (1%; $\kappa = .66$, agreement = 99%), while others provided broader positive feedback (e.g., "Using the Flourish app has helped me to gain clarity and move forward on a path to recovery and evolution of myself with ease"). Such results indicate that participants engaged attentively with the intervention. (For more details, see Supplementary Materials, Section 7.)

Notwithstanding, future research should incorporate active comparison conditions, app engagement metrics, and conversation scripts to identify which components (e.g., conversational coaching, activities, reflective insights) most strongly predict which outcomes. Such insights will help clarify how generative AI systems can be designed to enhance well-being and how they can integrate into human support networks—such as student success coaches, peer mentors, or therapists—to create hybrid models of well-being support that combine human empathy with AI immediacy, scalability, and accessibility. Longitudinal studies should also examine whether sustained use leads to lasting improvements in well-being, academic retention, performance, and healthcare utilization.

In the context of the ongoing mental health crisis, this research shows that a purpose-built generative AI system—grounded in psychological and behavioral science and guided by a closed-loop personalization framework—can help individuals and communities cultivate resilience, belonging, and collective flourishing in a manner that is early, scalable to the population level, and stigma-free. Contributing to the continuum of care, it indicates that evidence-based support can be extended beyond the limits of traditional counseling capacity, boosting and buffering mental health before distress escalates to impairing levels.

## Methods

### Participants

A total of 486 undergraduate students from Chapman University, the University of Washington, and Foothill College were enrolled in this study. Participants were recruited through the SONA research participation system at Chapman University, and through psychology courses at the University of Washington and Foothill College. Study periods and demographic characteristics by institution are reported in Extended Data Table 1.

Once enrolled at Week 0, participants' completion rate of each survey were almost identical between conditions (81% vs. 80% in Week 2; 74% vs. 72% in Week 4; 74% vs 70% in Week 6), and there were no significant differences in demographic characteristics across experimental conditions (age: $p = .460$; sex: $p = .320$; institution: $p = .813$). Across both conditions, 75 (15.4%) completed only one timepoint, 51 (10.5%) completed two timepoints, 40 (8.2%) completed three timepoints, and 320 (65.8%) completed all four timepoints. The sample sizes at each wave are shown in Extended Data Figure 1.

### Procedure

Study procedures were reviewed and approved by the Institutional Review Boards at each institution.

Participants completed four surveys across six weeks during the Fall 2024 semester (Weeks 0, 2, 4, and 6). All participants received extra credit, either as course credit (University of Washington and Foothill College) or research participation credit (Chapman University), on a prorated basis, such that students earned partial credit for each completed survey and full credit for completing all timepoints.

Surveys were administered using Qualtrics, which also handled random assignment to the treatment or control condition. At the end of the Week 0 baseline survey, students randomly assigned to the treatment condition were told the following: "As part of this study, you'll be using a free, science-based well-being app called Flourish." They were then shown QR codes to download the app and instructed to "try to use Flourish at least two days per week," however they wanted. At each follow-up survey, participants self-reported their app engagement by referencing an activity log in the Flourish app and entering those numbers (i.e., total number of active days and activities completed) into the survey. Given the complexities of coordinating a multi-institutional trial and securing multiple IRB approvals, we adopted a conservative approach to participant privacy, thus electing not to collect backend user engagement data and instead assessed app engagement via self-reported measures.

As a manipulation check, we examined behavioral engagement with the Flourish app. On average, students self-reported using the app 3.49 days per week ($SD = 3.46$) and completed 3.39 activities per week ($SD = 2.52$). Engagement was similar across the intervention period: students used the app for an average of 5.85 days ($SD = 4.23$) during Weeks 0–2, 4.58 days ($SD = 4.31$) during Weeks 2–4, and 5.46 days ($SD = 4.92$) during Weeks 4–6. Most students used the app at least once in each two-week interval (Week 0–2: 96%; Weeks 2–4: 78%; Weeks 4–6: 76%), and

72% engaged at least once during every interval. Overall, 43% of students met the recommended minimum of ≥2 days of use per week across all periods.

All analyses followed an intent-to-treat approach, including all participants regardless of engagement level. Longitudinal linear mixed-effects models were implemented in R (version 4.2.3) via the lmerTest package[41]. Time was zero-centered—the four timepoints were coded as -1.5, -0.5, 0.5, and 1.5 for Week 0, Week 2, Week 4, and Week 6, respectively.

**The Flourish App**

The Flourish app (https://www.myflourish.ai/) was developed for iOS and Android using Flutter and operates on a cloud-based backend hosted on Amazon Web Services (AWS).

*AI Architecture.* As an AI-native mobile application, the Flourish app's modular architecture integrates several interconnected AI services powered by several dozens of LLM prompts and workflows.

The Conversational Agent manages user-Sunnie dialogues within a chatbot interface and incorporates an advanced long-term memory system, multimodal interaction capabilities (including text, voice, and image), in-chat resource recommendation, and tool-calling functions (such as surfacing personalized insight summaries or launching relevant well-being activities directly within the chat). It is also connected to an automatic notecard generator that summarizes key takeaways, actionable tips, and notable insights after each conversation. The Personalized Recommendation Engine leverages LLMs to provide science-based well-being activity suggestions tailored to each user's moods, contexts, and engagement history. The Activity Library hosts a collection of pre-developed well-being activities offered either as AI-guided chat experiences such as "Three Good Things," or as non-AI interactive activities such as "Breathing with Sunnie," which features calming music, animation, haptic feedback, and a timer. A Habit-Building and Gamification System rewards users with AI-classified badges based on the PERMA model of well-being[24] for completing chats, activities, or journal entries, and tracks engagement through a continuously increasing streak counter. An opt-in Push Notification Service delivers pre-drafted or AI-generated daily reminders, positive affirmations, and check-in messages to encourage consistent engagement. A dedicated Crisis Service continuously scans for high-risk language and activates escalation protocols during the conversation. The Analytics and Insights module aggregates user data to produce weekly summaries. Finally, authentication and user management are handled through Firebase, and the infrastructure is orchestrated with Terraform, supported by continuous integration and deployment (CI/CD) pipelines powered by GitLab. At the time of this research, all conversational and recommendation prompts were powered by the OpenAI API, using either GPT-4o or GPT-4o-mini.

*Conversation Design and Validation.* Sunnie's chat behaviour is designed to balance listening, reflection, and gentle guidance, leveraging methods commonly found across psychological, affective, and behavioural sciences. For example, when users experience positive events, Sunnie guides them to savour and deepen their positive emotions by noticing, amplifying, sharing the experience with others, or expressing appreciation to others. When users face challenges or distressing experiences, Sunnie draws on evidence-based strategies to foster resilience and self-compassion, which may include techniques such as building emotional awareness, practicing cognitive reappraisal, engaging in expressive writing, recognizing and reframing unhelpful thoughts, engaging in mindfulness and relaxation exercises, affirming personal values, and taking actions aligned with personal goals. When appropriate, Sunnie also promotes social connection and help-seeking, guiding users toward constructive communication, conflict resolution, and professional support when needed. All prompts were developed by psychologists and iteratively refined through extensive testing, red-teaming, and collaboration with mental health researchers and clinicians with diverse training backgrounds prior to field deployment. This process involved systematic evaluation of tone calibration, guardrail performance, and adherence to evidence-based content, ensuring that responses were natural, insightful, empathic, safe, and trustworthy.

***Monitoring and Quality Assurance***. The AI-first, human-in-the-loop continuous monitoring pipeline tracks multiple layers of performance and safety data, including conversation-level metrics (e.g., user-rated conversation quality, AI-generated mental health risk assessments), individual-level metrics (e.g., mental health profiles and ratings), and system-level metrics (e.g., latency, response compliance, and escalation triggers). The conversation-level feedback is fed back to Sunnie to refine and personalize its conversational style for each user, while mental health assessments and trigger data are used for quality monitoring and assurance by the human review team. Reviewers provide targeted feedback to continuously improve the AI system, ensuring accuracy, clinical safety, and emotional resonance. Importantly, no user data is ever used to train the underlying Large Language Models (LLMs).

***Safety, Privacy, and Security.*** In addition to the OpenAI API's built-in guardrails, our purpose-built crisis identification and safety protocol operates automatically in real time. When potential risk statements are detected, the system triggers a just-in-time safety check to assess user safety and determine the appropriate course of action. When warranted, it displays relevant national resources (e.g., the 988 Suicide and Crisis Lifeline in the U.S.) or local community resources retrieved from our database using Retrieval-Augmented Generation (RAG), and generates a deidentified flag for human review.

Flourish is a confidential well-being resource. No personally identifiable information is shared with partner institutions, except in cases of imminent risk. All data are encrypted in transit and at rest using industry-standard protocols and stored securely in compliance with HIPAA (Health Insurance Portability and Accountability Act), GDPR (General Data Protection Regulation), and CCPA (California Consumer Privacy Act) requirements. Data are never sold or shared with external commercial entities. Access to information is restricted to authorized professionals working under strict data protection and ethical oversight, using secure, VPN-protected systems.

No medical advice or diagnosis was provided. No personally identifying information was stored for research purposes; all conversations were deidentified prior to human review, which is conducted solely to ensure AI safety, quality assurance, and program evaluation. Flagged conversations were reviewed by trained staff under the supervision of a clinical psychologist, following a predefined safety protocol.

## Measures

We used previously validated psychological instruments and organized our outcome variables into three domains: (1) emotional well-being, assessed using the Subcomponents of Affect Scale with positive and negative affect subscales[37]; (2) social well-being, assessed using measures of loneliness (UCLA 3-item Loneliness Scale[42]), campus belonging (Single-item from Perceived Cohesion Scale[43]), and felt closeness to one's university (Inclusion of Other in the Self Scale[44]); and (3) overall well-being, assessed using measures of resilience (Brief Resilience Scale[45]), mindfulness (Mindful Attention Awareness Scale[46]), and flourishing (Flourishing Scale[47]).

An exploratory factor analysis of baseline measures confirmed a three-factor structure consistent with this theoretical organization into emotional, social, and overall well-being domains, with minor expected overlap among constructs. Full factor loadings and details are provided in Supplementary Materials, Section 1, and detailed descriptions of all measures, subscales, and administration schedules are provided in Extended Data Table 2.

To capture more granular, temporal shifts in emotional and social well-being, we selected one construct from each domain—positive/negative affect and loneliness—to assess at every timepoint (Week 0, 2, 4, 6). All other measures were collected only at baseline (Week 0) and post-intervention (Week 6) to minimize participant burden while still detecting meaningful pre-to-post change.

For completeness we also assessed several deficit-oriented measures, including depression, anxiety, and stress, as well as academic self-efficacy. Although these additional measures did not yield significant interaction effects, we report full details for these measures and their results in Supplementary Materials, Sections 2 and 3.

## Open Science Practices

All study materials, data, and analysis code have been made publicly available here: https://github.com/Ethical-Intelligence-Lab/strengths_based_AI. Data were analyzed using R, version 4.2.3[48] and the package *ggplot*, version 3.5.0[49]. Our hypotheses were pre-registered at the following link: https://aspredicted.org/yrw2-szwp.pdf

**Extended Data Figure 1**

*Study Timeline, Sample Sizes, and Measures*

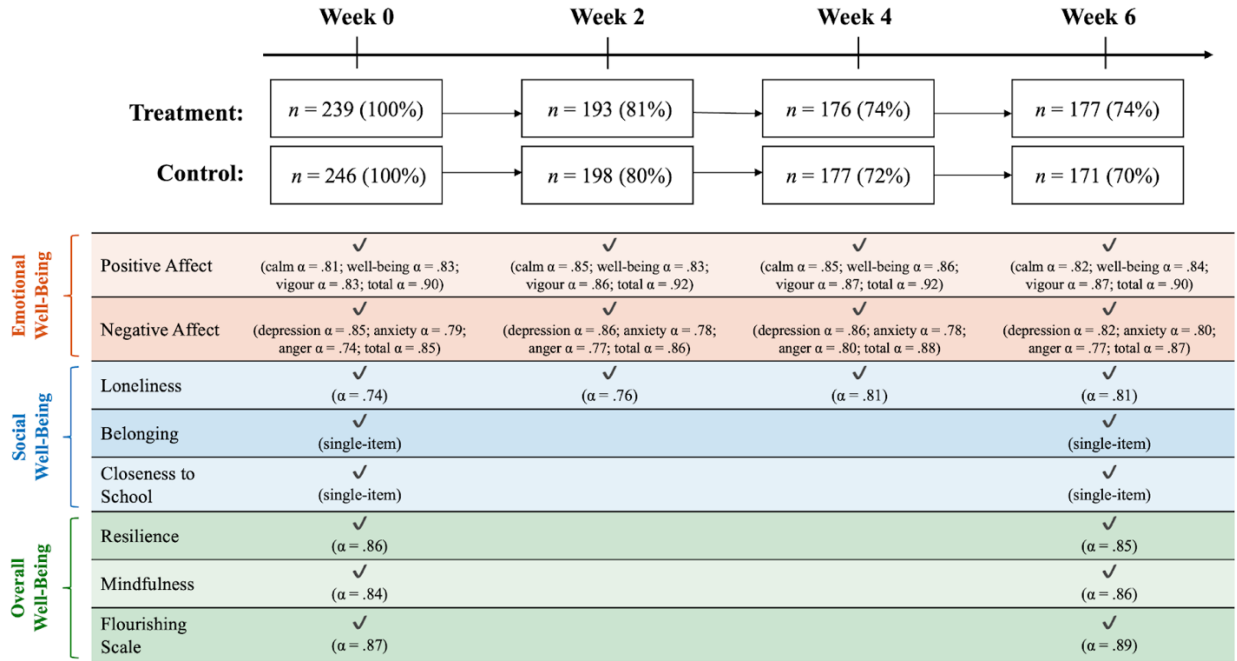

*Note. n*s reflect the number of students who completed the survey at each timepoint (i.e., available cases). Percentages in parentheses show retention relative to baseline sample recruited at Week 0. All analyses followed an intent-to-treat approach; participants who skipped a timepoint remained in the dataset for other timepoints.
Check marks denote that the measure was administered at that timepoint; blank cells indicate it was not. Cronbach's alpha (α) values are reported beneath each check mark where applicable.

**Extended Data Table 1**

*Study Period and Sample Demographics by Institution*

| Characteristic | Institution | | |
| --- | --- | --- | --- |
| | Chapman University | University of Washington | Foothill College |
| Study Period | 10/2/24 – 12/3/24 | 9/26/24 – 11/20/24 | 10/1/24 – 11/21/24 |
| Total Sample Size | $n = 245$ | $n = 175$ | $n = 66$ |
| Sex | 80% female; 20% male | 79% female; 19% male; 2% NA | 73% female; 27% male |
| Age (years) | $M = 19.75$, $SD = 1.99$ | $M = 19.74$, $SD = 2.25$ | $M = 24.20$, $SD = 8.71$ |
| Racial/Ethnic Composition | 53% White/European American; 10% Asian/Asian American; 20% mixed race; 10% Hispanic/Latin(x); 2% Black/African American; 3% Middle Eastern/Arab American; <1% American Indian; <1% Native Hawaiian/Pacific Islander; 2% other | 38% White/European American; 29% Asian/Asian American; 16% mixed race; 5% Hispanic/Latin(x); 6% Black/African American; 1% American Indian; 3% other | 24% White/European American; 18% Asian/Asian American; 15% mixed race; 23% Hispanic/Latin(x); 9% Black/African American; 3% Middle Eastern/Arab American; 2% Native Hawaiian/Pacific Islander; 6% other |
| Subjective SES (1-5) | $M = 3.31$, $SD = 1.15$ | $M = 3.32$, $SD = 1.11$ | $M = 3.23$, $SD = 1.25$ |
| International Students | 2.0% of sample | 9.9% of sample | 16.7% of sample |

# Extended Data Table 2

*Summary of Measures*

| Measure | Description | Items / Subscales | Scale Range | Timepoints Administered |
|---|---|---|---|---|
| **Positive and Negative Affect** | Subcomponents of Affect Scale (SAS; Jenkins et al., 2023); 18 adjectives (9 positive, 9 negative) | Positive: Calm (calm, at ease, relaxed), Well-being (happy, cheerful, pleased), Vigour (full of pep, lively, energetic); Negative: Depression (sad, unhappy, depressed), Anxiety (on edge, tense, nervous), Anger (hostile, angry, resentful) | 0 (not at all accurate) to 4 (extremely accurate); Composite scores: 0–36 (positive/negative) | All four timepoints |
| **Loneliness** | UCLA 3-item Loneliness Scale (Russell et al., 1980) | "Lack companionship", "Feel left out", "Feel isolated" | 1 (hardly ever) to 3 (often); Total: 3–9 | All four timepoints |
| **Belonging** | Single-item from Perceived Cohesion Scale (Bollen & Hoyle, 1990) | "I see myself as a part of the campus community" | 0 (strongly disagree) to 10 (strongly agree) | Beginning and end of study |
| **Closeness to School** | Single-item Inclusion of Other in the Self (IOS) Scale (Aron et al., 1992) | Visual selection of 1–7 overlapping circles | 1 (least overlap) to 7 (most overlap) | Beginning and end of study |
| **Resilience** | 6-item Brief Resilience Scale (Smith et al., 2008) | E.g., "I tend to bounce back quickly after hard times", "I have a hard time making it through stressful events" (reverse-coded) | 1 (strongly disagree) to 5 (strongly agree); Total: 6–30 | Beginning and end of study |
| **Mindfulness** | 5-item state version of Mindful Attention Awareness Scale (MAAS; Brown & Ryan, 2003) | E.g., "I was finding it difficult to stay focused", "I was doing things automatically" | 0 (not at all) to 6 (very much); Total: 0–30 (reverse coded so higher = more mindfulness) | Beginning and end of study |
| **Flourishing** | 8-item Flourishing Scale (Diener et al., 2010) | E.g., "I lead a purposeful and meaningful life", "My social relationships are supportive and rewarding" | 1 (strongly disagree) to 7 (strongly agree); Total: 8–56 | Beginning and end of study |

# Extended Data Table 3

*Means and Standard Deviations of Study Variables by Time and Condition*

| Measure | Condition | T1 (Baseline), M (SD) | T2 (Week 2), M (SD) | T3 (Week 4), M (SD) | T4 (Week 6), M (SD) |
|---|---|---|---|---|---|
| ***Emotional Well-Being*** | | | | | |
| Positive Affect | Control | 18.48 (6.58) | 17.35 (6.85) | 17.43 (7.30) | 17.21 (6.98) |
| | Treatment | 18.93 (6.90) | 18.51 (6.65) | 19.11 (6.76) | 19.38 (6.21) |
| *Calm* | Control | 5.54 (2.44) | 5.36 (2.57) | 5.47 (2.58) | 5.44 (2.58) |
| | Treatment | 5.72 (2.69) | 5.79 (2.59) | 6.05 (2.66) | 6.28 (2.41) |
| *Well-being* | Control | 7.03 (2.46) | 6.70 (2.47) | 6.57 (2.73) | 6.51 (2.56) |
| | Treatment | 7.24 (2.53) | 6.85 (2.35) | 7.16 (2.43) | 7.25 (2.34) |
| *Vigour* | Control | 5.91 (2.76) | 5.30 (2.69) | 5.39 (2.92) | 5.26 (2.88) |
| | Treatment | 5.97 (2.67) | 5.88 (2.61) | 5.92 (2.61) | 5.88 (2.59) |
| Negative Affect | Control | 13.35 (6.72) | 12.48 (6.54) | 12.16 (6.65) | 12.85 (7.05) |
| | Treatment | 12.84 (6.83) | 12.13 (6.61) | 11.84 (7.18) | 12.29 (6.62) |
| *Depression* | Control | 4.37 (2.99) | 4.07 (2.89) | 3.98 (2.96) | 4.11 (3.01) |
| | Treatment | 3.99 (3.05) | 3.75 (2.79) | 3.66 (2.99) | 3.82 (2.65) |
| *Anxiety* | Control | 6.48 (3.00) | 6.06 (2.86) | 5.85 (2.79) | 6.22 (2.90) |
| | Treatment | 6.21 (2.89) | 5.78 (2.78) | 5.49 (2.67) | 5.80 (2.86) |
| *Anger* | Control | 2.50 (2.29) | 2.35 (2.24) | 2.33 (2.36) | 2.52 (2.58) |
| | Treatment | 2.64 (2.56) | 2.60 (2.64) | 2.69 (2.85) | 2.70 (2.61) |
| ***Social Well-Being*** | | | | | |
| Loneliness | Control | 5.60 (1.64) | 5.28 (1.60) | 5.16 (1.70) | 5.40 (1.71) |
| | Treatment | 5.58 (1.65) | 5.08 (1.58) | 4.98 (1.66) | 5.10 (1.67) |
| Belonging | Control | 5.61 (2.07) | — | — | 5.68 (2.18) |
| | Treatment | 5.64 (2.22) | — | — | 6.11 (2.14) |
| Closeness to School | Control | 3.24 (1.40) | — | — | 3.31 (1.42) |
| | Treatment | 3.25 (1.39) | — | — | 3.52 (1.44) |
| ***Overall Well-Being*** | | | | | |
| Resilience | Control | 19.04 (4.61) | — | — | 18.73 (4.56) |
| | Treatment | 18.67 (4.48) | — | — | 19.31 (4.22) |
| Mindfulness | Control | 15.41 (5.82) | — | — | 13.67 (6.05) |
| | Treatment | 15.23 (6.12) | — | — | 14.99 (6.11) |
| Flourishing | Control | 44.97 (6.58) | — | — | 44.34 (6.62) |
| | Treatment | 44.59 (6.45) | — | — | 44.92 (6.59) |

*Note.* Dashes indicate measures not administered at that timepoint.